\documentclass{INTERSPEECH2023}

\interspeechcameraready 

\usepackage{adjustbox}

\usepackage{hyperref} 

\title{DeCoR: Defy Knowledge Forgetting by Predicting Earlier Audio Codes}
\name{Xilin Jiang, Yinghao Aaron Li, Nima Mesgarani}
\address{
    Department of Electrical Engineering, Columbia University, USA
}
\email{xj2289@columbia.edu, yl4579@columbia.edu, nima@ee.columbia.edu}

\begin{document}

\maketitle
 
\begin{abstract}

Lifelong audio feature extraction involves learning new sound classes incrementally, which is essential for adapting to new data distributions over time. However, optimizing the model only on new data can lead to catastrophic forgetting of previously learned tasks, which undermines the model's ability to perform well over the long term. This paper introduces a new approach to continual audio representation learning called DeCoR. Unlike other methods that store previous data, features, or models, DeCoR indirectly distills knowledge from an earlier model to the latest by predicting quantization indices from a delayed codebook. We demonstrate that DeCoR improves acoustic scene classification accuracy and integrates well with continual self-supervised representation learning. Our approach introduces minimal storage and computation overhead, making it a lightweight and efficient solution for continual learning.

\end{abstract}

\begin{figure*}[!th]
    \centering
    \includegraphics[scale=0.5]{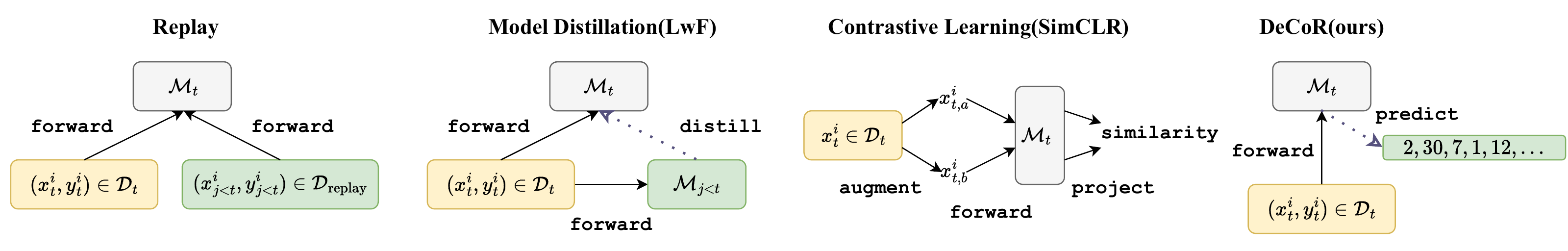}
    \caption{Comparison of DeCoR with other continual learning methods. The arrows indicate computation, and the green boxes indicate extra storage. We can observe that DeCoR is more efficient in terms of both computation and storage. Replay-based methods require storing and training on the audio rehearsal. Model knowledge distillation requires additional space and computation to store and forward one or more past model checkpoints. Contrastive learning demands training on multiple augmented views of the same audio. In contrast, DeCoR only stores and predicts one quantization index per audio.
    }
    \label{fig:1}
\end{figure*}

\section{Introduction}

Deep neural networks have significantly improved various audio processing tasks such as event detection \cite{PANN}, speech recognition \cite{SpeechStew}, and source separation \cite{SepIt}. However, most works assume an offline setting, where the entire dataset is available at the start of training. This assumption is unrealistic because new sound patterns, speaker accents, and background noises continuously emerge as time passes or as location changes. In practice, keeping all previous data and retraining the model can be either impossible or too expensive due to data privacy, hardware storage, and computation cost. Fine-tuning the model directly on new data can adapt it to the latest data distribution but may come at a cost of forget previously learned knowledge, resulting in what is known as \textit{catastrophic forgetting} \cite{cl_survey}.

Continual learning or lifelong learning aims to overcome forgetting and enhance long-term performance for a series of tasks \cite{cl_survey, cl_survey2}. In each task, the model can only access the current data with little or no access to the previous data. A variety of methods have been proposed under this setting, which can be categorized into three types: \textit{regularization}, \textit{replay}, and \textit{parameter isolation} \cite{cl_survey}. Recently, self-supervised contrastive methods have been shown to acquire more task-persistent knowledge in both image \cite{cssl} and audio \cite{audio_cssl} domains. Although these methods significantly reduce forgetting without retraining on all data, the computation and storage overhead is still significant. For example, \cite{lwf, multiEncoderDistill} store and forward the last or all previous model(s) to teach the current model earlier task knowledge. Replay methods \cite{icarl, A-GEM} store selected exemplars from previous tasks and retrain the model on them. Progressive model \cite{PNN, cl_lst} gradually enlarges the model size after each task. Contrastive learning like SimCLR \cite{SimCLR} generates at least two augmented views of each audio instance, resulting in twice the computation. Thus, there is a need for more efficient continual learning methods to tackle the above overhead.

Here, we present a new method for efficient continual learning called \textbf{De}layed \textbf{Co}debook \textbf{R}egularizier (DeCoR). Unlike other regularization-based methods, DeCoR does not require extra storage for the teacher model. Instead, it only stores a single quantization index for each audio in the new dataset. The index corresponds to the nearest code in the feature codebook generated by the model before training on the new dataset. Once these indices are obtained, the codebook can be freed from memory. By predicting the delayed indices from the current features, DeCoR indirectly transfers the knowledge from the previous model to the current one. We expect the DeCoR-regularized model to perform better on earlier tasks with minimal additional computation and storage. The comparison between DeCoR and other continual learning methods is shown in Figure \ref{fig:1}. DeCoR is a straightforward method that can be applied to various audio processing tasks. 

In this study, we evaluate the effectiveness of DeCoR on class-incremental sound classification \cite{ae_gr} from features trained with or without DeCoR. In addition, we integrate DeCoR into the continual self-supervised representation learning framework \cite{audio_cssl} to investigate its impact on self-supervised knowledge transfer. Our experiments demonstrate that DeCoR outperforms simple finetuning or knowledge distillation from the previous model for both supervised and self-supervised settings with a higher long-term accuracy and lower forgetting.

\section{Related work}
Traditional replay and knowledge distillation methods require storage of some training data or the teacher models to mitigate forgetting \cite{lwf, multiEncoderDistill, icarl, A-GEM}. A straightforward alternative to reduce storage is to compress input data or model features. For example, AQM \cite{aqm} learns adaptive multi-level image compression and reconstruction along with continual classification to maximize the number of samples given a fixed replay budget. IB-DRR \cite{IB-DRR} further proposes a two-step compression and maintains the replay as codes. Another direction is to use codes as training targets, such as \cite{spq} which replaces the projection layer in contrastive learning with a quantization module and tries to maximize the similarity between the continuous representation of one view and the quantized representation of the other. 

In contrast to these methods, DeCoR only keeps and predicts quantization indices. The regularization loss is calculated between the predicted and actual indices, not between the estimated features and codes. Code index prediction has already been adopted as an SSL objective in HuBERT \cite{hubert} and WavLM \cite{wavlm}, where masked cluster indices from K-means of either MFCC or learned representations are predicted. This forces the model to learn contextual information. Later, MVQ-KD \cite{multi_index} proposes to do knowledge distillation from HuBERT to conformer \cite{conformer} by predicting HuBERT code indices which achieves comparable performance but with much fewer data, model size, and training time. Our method is similar to MVQ-KD but targets continual learning rather than model compression. Our codebook is manually constructed from the same but earlier model instead of another larger teacher model.

\section{Method}

\subsection{Problem definition}
We proceed with the standard continual learning setup, where we have an ordered sequence of $T$ tasks $[\mathcal{T}_1, \mathcal{T}_1, ..., \mathcal{T}_T]$ with task boundaries given. Each task  $\mathcal{T}_t$ is associated with a dataset $\mathcal{D}_t = \{(x_t^{(i)}, y_t^{(i)})\}$ with both different audios and their corresponding labels. We aim to train an audio feature extractor $\mathcal{M}$ on this sequence with only access to the current data but working well on all previous tasks. We evaluate $\mathcal{M}_t$, the model at the end of $\mathcal{T}_t$,  by sound classification accuracy of all seen tasks $j, j \leq t$. We define $A_{t, j}$ to be the accuracy of $\mathcal{M}_t$ on $\mathcal{D}_j$, and we aim to maximize the average accuracy on seen classes $A_{t}$ while minimizing the largest forgetting $F_{t}$ from the historical best for every $t$, especially the last task $t=T$:
\begin{align}
  A_{t} = \frac{1}{t} \sum_{j=1}^{t} A_{t, j}, \hspace{4pt}
  F_{t} = \frac{1}{t-1} \sum_{j=1}^{t-1} \underset{1 \leq \tau \leq t}{\max} (A_{\tau, j} - A_{t, j})
\end{align}

\subsection{Delayed Codebook Regularizer}
We can consider the Delayed Codebook Regularizer (DeCoR) as a simplified version of model distillation methods such as LwF \cite{lwf}. In LwF, knowledge is transferred from the previous to the current model by comparing the output probabilities of both models for the same input. However, with DeCoR, we avoid the need to store and forward the teacher model by using a selected set of teacher features, which we call the delayed codebook. To regularize the current model, we pair each training audio with the closest code from the codebook. This reduces the problem of model-to-model to codebook-to-model distillation.

DeCoR uses index prediction \cite{hubert, wavlm, multi_index} as its distillation objective, which involves predicting the index of the closest code in the delayed codebook for each training example. While other options, such as minimizing the distance between features and codes, are available, DeCoR opts for index prediction since it further reduces storage requirements by only keeping the indices as pseudo training labels. Additionally, this objective is compatible with the model's original training objective, including contrastive loss, as demonstrated in our experiments.

\begin{figure*}[!th]
    \centering
    \includegraphics[scale=1.07]{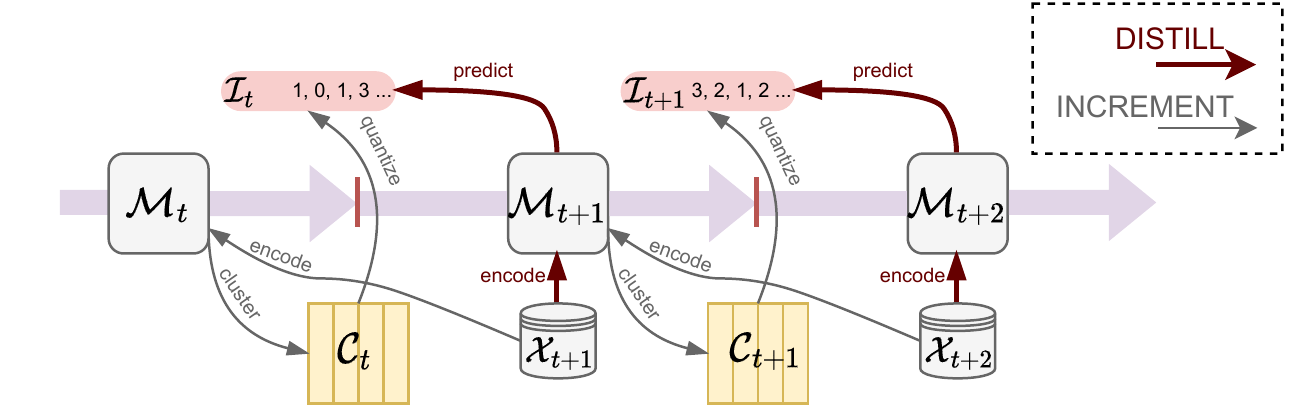}
    \caption{Graphical illustration of how DeCoR works with irrelevant model components omitted for clarity. The gray arrows correspond to actions taken during the \textit{INCREMENT} step at the task boundary, where new task data is encoded, clustered, and indexed using the delayed codebook. The red arrows correspond to actions taken during the \textit{DISTILL} step throughout the task, where the model is trained to predict the indices assigned earlier in the \textit{INCREMENT} step. Notably, there is no direct connection between the current and previous models, and knowledge is distilled solely through index prediction. Past models and codebooks are depicted for illustration purposes, with only quantization indices being stored.
    }
    \label{fig:2}
\end{figure*}

The process of building the delayed codebook and regularizing the model is conducted in an alternating fashion, as illustrated in Figure \ref{fig:2}. We refer to the step of constructing the codebook $\mathcal{C}_{t}$ and assigning indices $\mathcal{I}_{t}$ at the task boundary between $\mathcal{T}_{t}$ and $\mathcal{T}_{t+1}$ as \textit{INCREMENT}. On the other hand, we refer to the step of training model  $\mathcal{M}_{t+1}$ with indices $\mathcal{I}_{t}$ throughout a new task $\mathcal{T}_{t+1}$ as \textit{DISTILL}. In the \textit{INCREMENT} step, the model is fixed while the latest knowledge is encoded in the delayed codebook. During the \textit{DISTILL} step, the delayed knowledge (represented by $\mathcal{C}_t$ and $\mathcal{I}_t$) compared to $\mathcal{M}_{t+1}$ is transferred back to the model by predicting the indices assigned in \textit{INCREMENT}.

\subsubsection{INCREMENT}
The \textit{INCREMENT} step is executed at the task boundary, just before the model $\mathcal{M}_{t}$ is updated on the new task $\mathcal{T}_{t+1}$. To construct the delayed codebook $\mathcal{C}_{t}$, we encode the new audio data $\mathcal{X}_{t+1}$ using the fixed parameters of $\mathcal{M}_{t}$ to obtain features $\Phi_{t}$. Subsequently, we apply K-means clustering to $\Phi_{t}$, with $K$ being the desired codebook size, to create the codebook $\mathcal{C}_{t}$. At the final iteration of K-means, we obtain the cluster assignment (i.e., the nearest code index) $\mathbf{i}_t^i \in \mathcal{I}_{t}$ for each encoded feature $\phi^{i}_{t} \in \Phi_{t}$. At this point, we can discard the codebook $\mathcal{C}_{t}$ and only retain the quantization indices $\mathcal{I}_{t}$. The equations that define the \textit{INCREMENT} step are provided below:
\begin{align}
  \Phi_{t} &= \mathcal{M}_t(\mathcal{X}_{t+1}) \nonumber \\
  \mathcal{C}_{t},  \mathcal{I}_{t} &= \text{K-means}(\Phi_{t})
\end{align}
Recall that K-means updates the clusters to minimize the $\mathcal{L}_2$ distance between each feature $\phi_{t+1}^{i}$ and its nearest code $\mathcal{C}_{t}[\mathbf{i}_t^i]$:
\begin{align}
    \underset{\mathcal{C}_{t}}{\min} \sum_i || \phi_{t}^{i} - \mathcal{C}_{t}[\mathbf{i}_t^i]||^2, \hspace{4pt} \ \mathbf{i}_t^i = \underset{\mathbf{i}=0, ..., K-1}{\text{argmin}} || \phi_{t}^{i} - \mathcal{C}_{t}[\mathbf{i}]||^2
\end{align}

\subsubsection{DISTILL}
DeCoR distills the knowledge indirectly from the model $\mathcal{M}_{t}$ trained on previous tasks to the latest model $\mathcal{M}_{t+1}$ trained on the current task by letting the latter predict the quantization index of the former. We initialize an index prediction head $\mathcal{P}_{t+1}$ with the output dimension equal to the number of codes $K$. $\mathcal{P}_{t+1}$ and $\mathcal{M}_{t+1}$ are optimized jointly using the cross entropy loss, with the quantization index serving as the pseudo label.
\begin{align}
    \phi_{t+1}^{i} &= \mathcal{M}_{t+1}(x_{t+1}^i) \nonumber \\
     p_{t+1}^{i} &= \mathcal{P}_{t+1}(\phi_{t+1}^{i}) \nonumber \\
    \mathcal{L}_{pd}^i &= \text{CrossEntropy}(p_{t+1}^{i}, \mathbf{i}_t^i)
\end{align}

\subsubsection{Training objective}
We apply DeCoR to the acoustic scene classification task. The original loss for the task is cross entropy $\mathcal{L}_{ce}$ between the scene labels and the predicted scene probabilities from the sound classification head $\mathcal{H}_{t+1}$ on top of the encoder $\mathcal{M}_{t+1}$. After the first task, we initialize an index prediction head $\mathcal{P}_{t+1}$ and optimize the DeCoR regularization loss $\mathcal{L}_{pd}$ between the quantization indices and the predicted index probabilities. The combined loss for the subsequent task is:
\begin{align}
    \mathcal{L} = \mathcal{L}_{ce} + \lambda \mathcal{L}_{pd}
\end{align}

We also incorporate DeCoR with self-supervised learning (SSL) techniques like SimCLR \cite{SimCLR}. SimCLR trains on audios without labels. Therefore, we need to replace the classification head $\mathcal{H}_t$ with the projection head $\mathcal{F}_t$ and the supervised cross-entropy loss $\mathcal{L}_{ce}$ with NT-Xent loss $\mathcal{L}_{ssl}$. After these modifications, the final loss becomes:
\begin{align}
    \mathcal{L} = \mathcal{L}_{ssl} + \lambda \mathcal{L}_{pd}
\end{align}

\subsubsection{Overhead}
DeCoR requires an additional epoch, referred to as epoch 0, when entering a new task to \textit{INCREMENT} the codebook with the model weights frozen. However, Epoch 0 is insignificant compared to the full training up to tens or hundreds of epochs per task. In \textit{DISTILL} step, a shallow index predictor is added to the feature extractor for knowledge distillation. The size of the predictor is relatively small compared to the model backbone, and We explore the effect of the number of predictor layers in \textit{ablation}. 

Once each new audio is assigned to the nearest code index, we can release the codebook from memory and directly update the model without making a checkpoint copy. The extra storage for indexing each audio is minimal, one integer or $\log_2 K$ bits.

\begin{figure*}[!th]
    \centering
    \includegraphics[width=\linewidth]{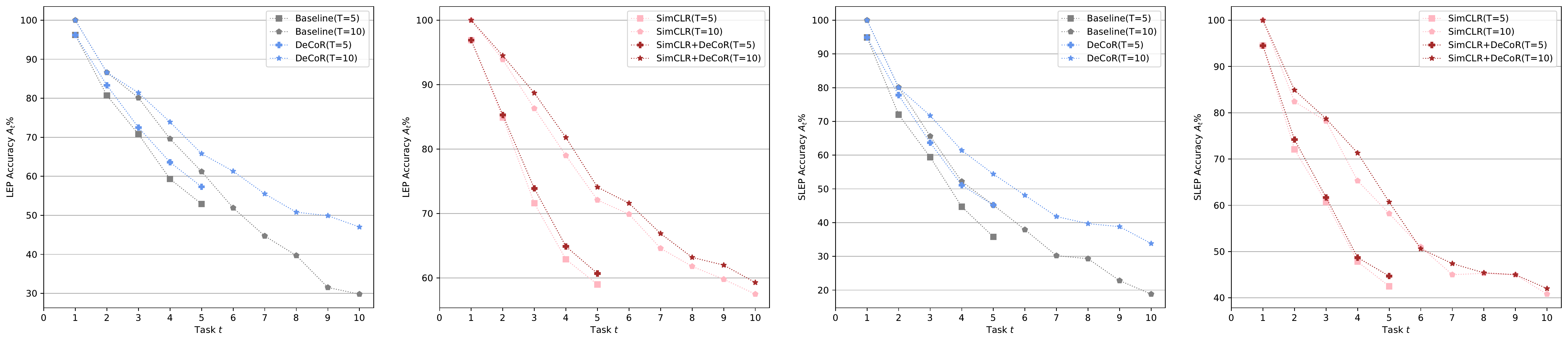}
    \caption{LEP and SLEP accuracy evaluated at the end of every task. The order from left to right is as follows: LEP for supervised training, LEP for self-supervised training, SLEP for supervised training, and SLEP for self-supervised training. We observe an improvement in $A_t$ with DeCoR starting from $t=2$. An exception is $t=2$ for 10-task supervised training, where the model only learns one class at $t=1$, making it too trivial to classify and resulting in the distilled knowledge from $t=1$ being useless.
    }
    \label{fig:3}
\end{figure*}

\section{Experiments}
\subsection{Dataset}
In our experiments, we use TAU Urban Acoustic Scenes 2019 dataset \cite{tau19} for incremental acoustic scene classification. The dataset has 1440 audio segments, each representing one of the 10 acoustic scenes recorded in multiple locations. We randomly split 10 classes into $T=5$ or $10$ tasks. The classes are evenly divided and disjoint for each task. For $T=5$, the model learns two new classes per task. For $T=10$, the model learns one new class per task.

\begin{table}[!t]
	\centering
	\caption{Final accuracy and forgetting for 5 tasks.\\} 
    \begin{adjustbox}{width=\columnwidth,center}
    \begin{tabular}{c|c|c|c|c|c}
    & \multicolumn{2}{c}{LEP} & \multicolumn{2}{c}{SLEP} \\
    Method & $A_5 \uparrow $ & $F_5 \downarrow $ & $A_5 \uparrow $ & $F_5 \uparrow$ & Storage $\downarrow$ \\
    \hline\hline
    Baseline & 52.9 & 23.6 & 35.8 & 36.1 & 0 B \\
    LwF & 54.0 & 19.1 & 40.3 & 30.5 & 302 MB \\
    \textbf{DeCoR} & 57.3 & 17.1 & \textbf{45.2} & 23.4 & 4.08 KB \\
    \hline
    SimCLR & 59.0 & 18.8 & 42.5 & 24.7 & 0 B \\
    SimCLR + LwF & 59.5 & 17.1 & 41.4 & 28.1 & 302 MB \\
    \textbf{SimCLR + DeCoR} &\textbf{60.7} & \textbf{15.5} & 44.7 & \textbf{21.8} & 4.08 KB \\
    \hline
    \end{tabular}
    \end{adjustbox}
    \label{t1}
\end{table}
\begin{table}[!t]

	\centering
	\caption{Final accuracy and forgetting for 10 tasks.\\} 
\begin{adjustbox}{width=\columnwidth,center}
    \begin{tabular}{c|c|c|c|c|c}
    & \multicolumn{2}{c}{LEP} & \multicolumn{2}{c}{SLEP} \\
    Method & $A_{10} \uparrow $ & $F_{10} \downarrow $ & $A_{10} \uparrow $ & $F_{10} \downarrow $ & Storage  $\downarrow$ \\
    \hline\hline
    Baseline & 29.8 & 32.7 & 18.8 & 35.0 & 0 B \\
    LwF & 34.7 & 28.8 & 24.7 & 28.6 & 302 MB \\
    \textbf{DeCoR} & 47.0 & 20.5 & 33.8 & 21.8 & 2.26 KB \\
    \hline
    SimCLR & 57.5 & 17.3 & 40.8 & 20.4 & 0 B \\
    SimCLR + LwF & 57.7 & 18.3 & 41.2 & 23.4 & 302 MB \\
    \textbf{SimCLR + DeCoR} & \textbf{59.3} & \textbf{16.9} & \textbf{42.0} & \textbf{19.4} & 2.26 KB \\
    \hline
    \end{tabular}
    \end{adjustbox}

    \label{t2}
\end{table}

\subsection{Model training and evaluation}
Our audio encoder consists of the first 12 convolution layers of CNN14 \cite{PANN} with output dimension of 2048. To ensure a fair comparison of supervised and self-supervised representations, we follow the training and evaluation setups in \cite{audio_cssl}. In the continual learning step, we train the encoder $\mathcal{M}$ incrementally on the sequence of tasks using either the supervised classification head $\mathcal{H}$ with loss $\mathcal{L}_{ce}$ or the self-supervised SimCLR head $\mathcal{F}$ with loss $\mathcal{L}_{ssl}$, for 100 epochs. In the evaluation step, we take the pretrained $\mathcal{M}$, freeze its weights, and train a linear classifier on top with either all task data (Linear Evaluation Protocol or LEP) or 200 samples per task (Subset Linear Evaluation Protocol or SLEP), for another 50 epochs.

The evaluation step is identical for all experiments. In the continual learning step, the Baseline experiment finetunes the model with only $\mathcal{L}_{ce}$, the SimCLR experiments create two augmented views per audio segment, the LwF experiments add a distillation loss to minimize the KL divergence between the old and new classification probabilities or between the old and new SimCLR projections, and the DeCoR experiments use a $L = 2$ layer index predictor, codebook size $K = 32$, and loss weight $\lambda = 0.2$. SpecAugment \cite{specaugment} is used for all experiments, while no data augmentation is used for DeCoR's \textit{INCREMENT}. We follow the predefined data splits of TAU dataset. All methods are implemented using Speechbrain \cite{speechbrain} and are executed on an NVIDIA A40 GPU.

\subsection{Results}
The results are summarized in Table \ref{t1} and \ref{t2}, which show the final average seen accuracy $A_{T}$ and the largest forgetting $F_{T}$ for $T=5$ and $T=10$ tasks, respectively.  The extra storage is shown in the "Storage" column, and the number is averaged across tasks. Several observations can be made: (1). DeCoR improves the accuracy and reduces the forgetting for both 5 and 10 tasks on both supervised and SimCLR representations. (2). DeCoR outperforms model distillation LwF despite using much fewer resources. (3). DeCoR and SimCLR together produce stronger results and outperform the baseline LEP accuracy by around 8\%(T=5) and 30\%(T=10).

Figure \ref{fig:3} plots the LEP and SLEP $A_{t}$ at the end of each task with and without DeCoR in both supervised and SimCLR training. We observe a consistent improvement with DeCoR starting from the second task, with a significantly larger performance gain for larger $t$ in supervised training.

To summarize, our results demonstrate that predicting earlier audio codes is an effective regularization technique that can be used in both supervised and self-supervised training.

\begin{figure}[!h]
    \centering
    \includegraphics[width=\linewidth]{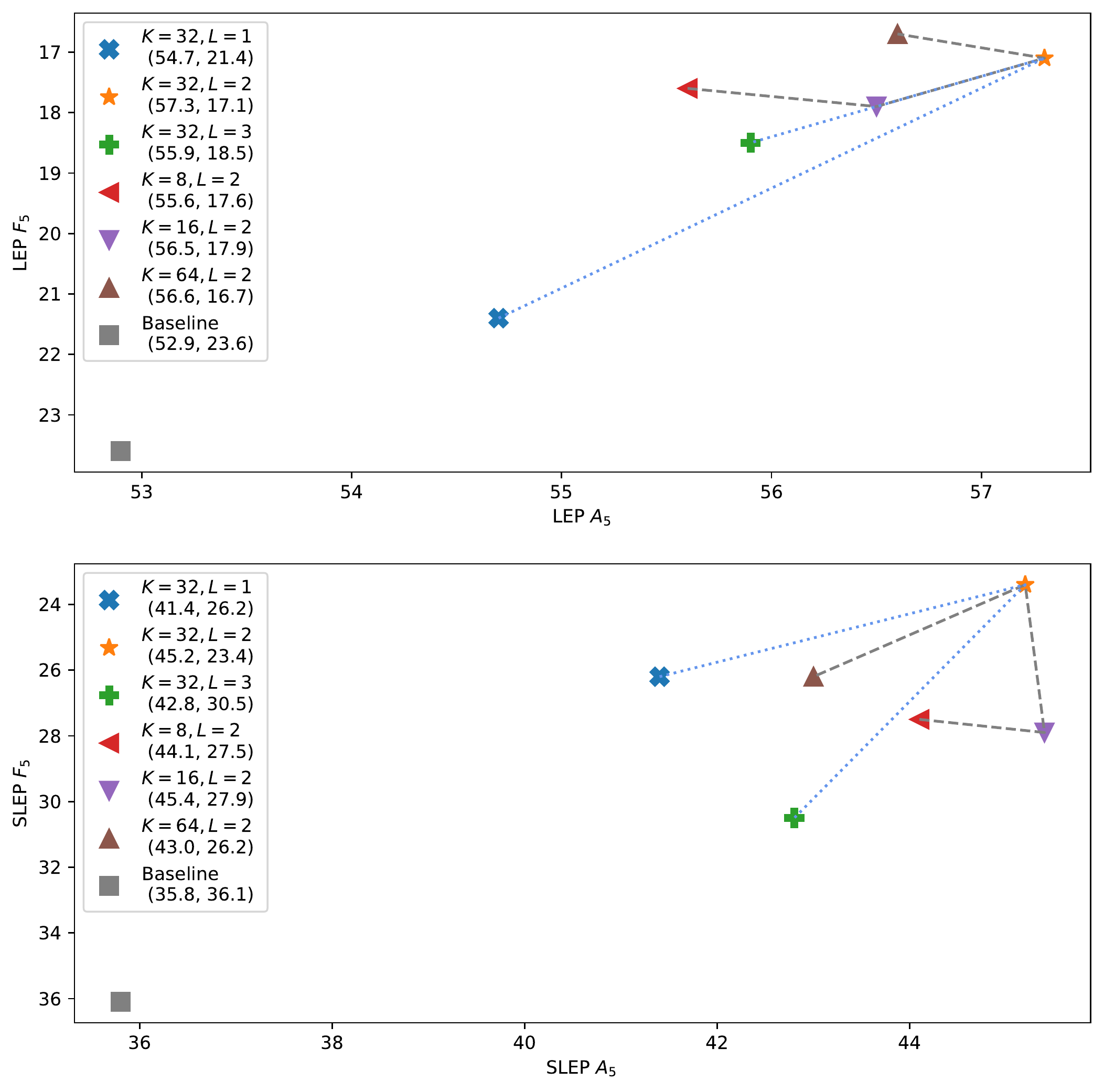}
    \caption{
    Impact of DeCoR codebook size $K$ and the predictor layer number $L$ to the final LEP (top) and SLEP (bottom) accuracy and forgetting for 5-task supervised training. All combinations result in better performance compared to the Baseline.
    }
    \label{fig:4}
\end{figure}

\subsection{Ablation}
The codebook size $K \in \{8, 16, 32, 64\}$ and the number of index prediction layers $L \in  \{1, 2, 3\}$ are tried for supervised $T=5$ experiments. The LEP and SLEP results are displayed in Figure \ref{fig:4}, with better performance being towards the top-right corner. The performance is improved from a linear ($L=1$) to a nonlinear ($L=2$) index predictor and from 8 codes to 32 codes. However, adding a third layer or increasing the number of codes does not lead to further improvement. We choose $K=32$ and $L=2$ in this paper. We hypothesize that the optimal choice of $K$ and $L$ depends on the feature dimension, the dataset size, and the downstream task, which leaves for future study.

\section{Conclusion}
In this work, we present a simple and efficient continual learning regularizer called DeCoR to reduce model knowledge forgetting. We apply DeCoR to both supervised and self-supervised continual representation learning and show experimentally that models trained with DeCoR achieve higher accuracy and lower forgetting on acoustic scene classification. In future research, we aim to extend the application of  DeCoR to other audio processing tasks, such as speaker identification and speech recognition, and adapt it to online learning problems without task boundaries.

\section{Acknowledgement}
This work was funded by the national institute of health
(NIH-NIDCD) and a grant from Marie-Josée and Henry R. Kravis. We used ChatGPT to improve the readability of the writing.

\newpage
\bibliographystyle{IEEEtran}
\bibliography{mybib}

\begin{thebibliography}{10}
\providecommand{\url}[1]{#1}
\csname url@samestyle\endcsname
\providecommand{\newblock}{\relax}
\providecommand{\bibinfo}[2]{#2}
\providecommand{\BIBentrySTDinterwordspacing}{\spaceskip=0pt\relax}
\providecommand{\BIBentryALTinterwordstretchfactor}{4}
\providecommand{\BIBentryALTinterwordspacing}{\spaceskip=\fontdimen2\font plus
\BIBentryALTinterwordstretchfactor\fontdimen3\font minus
  \fontdimen4\font\relax}
\providecommand{\BIBforeignlanguage}[2]{{%
\expandafter\ifx\csname l@#1\endcsname\relax
\typeout{** WARNING: IEEEtran.bst: No hyphenation pattern has been}%
\typeout{** loaded for the language `#1'. Using the pattern for}%
\typeout{** the default language instead.}%
\else
\language=\csname l@#1\endcsname
\fi
#2}}
\providecommand{\BIBdecl}{\relax}
\BIBdecl

\bibitem{PANN}
Q.~Kong, Y.~Cao, T.~Iqbal, Y.~Wang, W.~Wang, and M.~D. Plumbley, ``Panns:
  Large-scale pretrained audio neural networks for audio pattern recognition,''
  \emph{IEEE/ACM Transactions on Audio, Speech, and Language Processing},
  vol.~28, pp. 2880--2894, 2020.

\bibitem{SpeechStew}
\BIBentryALTinterwordspacing
W.~Chan, D.~Park, C.~Lee, Y.~Zhang, Q.~Le, and M.~Norouzi, ``Speechstew: Simply
  mix all available speech recognition data to train one large neural
  network,'' 2021. [Online]. Available: \url{https://arxiv.org/abs/2104.02133}
\BIBentrySTDinterwordspacing

\bibitem{SepIt}
S.~Lutati, E.~Nachmani, and L.~Wolf, ``Sepit: Approaching a single channel
  speech separation bound,'' in \emph{Interspeech}, 2022.

\bibitem{cl_survey}
M.~De~Lange, R.~Aljundi, M.~Masana, S.~Parisot, X.~Jia, A.~Leonardis,
  G.~Slabaugh, and T.~Tuytelaars, ``A continual learning survey: Defying
  forgetting in classification tasks,'' \emph{IEEE Transactions on Pattern
  Analysis and Machine Intelligence}, vol.~44, no.~7, pp. 3366--3385, 2022.

\bibitem{cl_survey2}
\BIBentryALTinterwordspacing
L.~Wang, X.~Zhang, H.~Su, and J.~Zhu, ``A comprehensive survey of continual
  learning: Theory, method and application,'' 2023. [Online]. Available:
  \url{https://arxiv.org/abs/2302.00487}
\BIBentrySTDinterwordspacing

\bibitem{cssl}
\BIBentryALTinterwordspacing
E.~Fini, V.~G.~T. da~Costa, X.~Alameda{-}Pineda, E.~Ricci, K.~Alahari, and
  J.~Mairal, ``Self-supervised models are continual learners,'' \emph{CoRR},
  vol. abs/2112.04215, 2021. [Online]. Available:
  \url{https://arxiv.org/abs/2112.04215}
\BIBentrySTDinterwordspacing

\bibitem{audio_cssl}
Z.~Wang, C.~Subakan, X.~Jiang, J.~Wu, E.~Tzinis, M.~Ravanelli, and
  P.~Smaragdis, ``Learning representations for new sound classes with continual
  self-supervised learning,'' \emph{IEEE Signal Processing Letters}, vol.~29,
  pp. 2607--2611, 2022.

\bibitem{lwf}
Z.~Li and D.~Hoiem, ``Learning without forgetting,'' \emph{IEEE Transactions on
  Pattern Analysis and Machine Intelligence}, vol.~40, pp. 2935--2947, 2016.

\bibitem{multiEncoderDistill}
\BIBentryALTinterwordspacing
Z.~Li, C.~Zhong, S.~Liu, R.~Wang, and W.-S. Zheng, ``Preserving earlier
  knowledge in continual learning with the help of all previous feature
  extractors,'' 2021. [Online]. Available:
  \url{https://arxiv.org/abs/2104.13614}
\BIBentrySTDinterwordspacing

\bibitem{icarl}
S.-A. Rebuffi, A.~Kolesnikov, G.~Sperl, and C.~H. Lampert, ``icarl: Incremental
  classifier and representation learning,'' \emph{2017 IEEE Conference on
  Computer Vision and Pattern Recognition (CVPR)}, pp. 5533--5542, 2016.

\bibitem{A-GEM}
A.~Chaudhry, M.~Ranzato, M.~Rohrbach, and M.~Elhoseiny, ``Efficient lifelong
  learning with a-gem,'' in \emph{ICLR}, 2019.

\bibitem{PNN}
\BIBentryALTinterwordspacing
A.~A. Rusu, N.~C. Rabinowitz, G.~Desjardins, H.~Soyer, J.~Kirkpatrick,
  K.~Kavukcuoglu, R.~Pascanu, and R.~Hadsell, ``Progressive neural networks,''
  \emph{CoRR}, vol. abs/1606.04671, 2016. [Online]. Available:
  \url{http://arxiv.org/abs/1606.04671}
\BIBentrySTDinterwordspacing

\bibitem{cl_lst}
\BIBentryALTinterwordspacing
X.~Guo, Y.~Tian, Q.~Xue, P.~Lampropoulos, S.~Eliuk, K.~Barner, and X.~Wang,
  ``Continual learning long short term memory,'' in \emph{Findings of the
  Association for Computational Linguistics: EMNLP 2020}.\hskip 1em plus 0.5em
  minus 0.4em\relax Online: Association for Computational Linguistics, Nov.
  2020, pp. 1817--1822. [Online]. Available:
  \url{https://aclanthology.org/2020.findings-emnlp.164}
\BIBentrySTDinterwordspacing

\bibitem{SimCLR}
\BIBentryALTinterwordspacing
T.~Chen, S.~Kornblith, M.~Norouzi, and G.~E. Hinton, ``A simple framework for
  contrastive learning of visual representations,'' in \emph{Proceedings of the
  37th International Conference on Machine Learning, {ICML} 2020, 13-18 July
  2020, Virtual Event}, ser. Proceedings of Machine Learning Research, vol.
  119.\hskip 1em plus 0.5em minus 0.4em\relax {PMLR}, 2020, pp. 1597--1607.
  [Online]. Available: \url{http://proceedings.mlr.press/v119/chen20j.html}
\BIBentrySTDinterwordspacing

\bibitem{ae_gr}
Z.~Wang, C.~Subakan, E.~Tzinis, P.~Smaragdis, and L.~Charlin, ``Continual
  learning of new sound classes using generative replay,'' in \emph{2019 IEEE
  Workshop on Applications of Signal Processing to Audio and Acoustics
  (WASPAA)}, 2019, pp. 308--312.

\bibitem{aqm}
L.~Caccia, E.~Belilovsky, M.~Caccia, and J.~Pineau, ``Online learned continual
  compression with adaptive quantization modules,'' \emph{Proceedings of the
  37th International Conference on Machine Learning}, 2020.

\bibitem{IB-DRR}
J.~Jiang, E.~Cetin, and O.~Celiktutan, ``Ib-drr - incremental learning with
  information-back discrete representation replay,'' in \emph{2021 IEEE/CVF
  Conference on Computer Vision and Pattern Recognition Workshops (CVPRW)},
  2021, pp. 3528--3537.

\bibitem{spq}
Y.~K. Jang and N.~I. Cho, ``Self-supervised product quantization for deep
  unsupervised image retrieval,'' in \emph{Proceedings of the International
  Conference on Computer Vision (ICCV)}, 2021.

\bibitem{hubert}
W.-N. Hsu, B.~Bolte, Y.-H.~H. Tsai, K.~Lakhotia, R.~Salakhutdinov, and
  A.~Mohamed, ``Hubert: Self-supervised speech representation learning by
  masked prediction of hidden units,'' \emph{IEEE/ACM Transactions on Audio,
  Speech, and Language Processing}, vol.~29, pp. 3451--3460, 2021.

\bibitem{wavlm}
\BIBentryALTinterwordspacing
S.~Chen, C.~Wang, Z.~Chen, Y.~Wu, S.~Liu, Z.~Chen, J.~Li, N.~Kanda,
  T.~Yoshioka, X.~Xiao, J.~Wu, L.~Zhou, S.~Ren, Y.~Qian, Y.~Qian, J.~Wu,
  M.~Zeng, X.~Yu, and F.~Wei, ``Wavlm: Large-scale self-supervised pre-training
  for full stack speech processing,'' \emph{{IEEE} J. Sel. Top. Signal
  Process.}, vol.~16, no.~6, pp. 1505--1518, 2022. [Online]. Available:
  \url{https://doi.org/10.1109/JSTSP.2022.3188113}
\BIBentrySTDinterwordspacing

\bibitem{multi_index}
\BIBentryALTinterwordspacing
L.~Guo, X.~Yang, Q.~Wang, Y.~Kong, Z.~Yao, F.~Cui, F.~Kuang, W.~Kang, L.~Lin,
  M.~Luo, P.~Zelasko, and D.~Povey, ``Predicting multi-codebook vector
  quantization indexes for knowledge distillation,'' 2022. [Online]. Available:
  \url{https://arxiv.org/abs/2211.00508}
\BIBentrySTDinterwordspacing

\bibitem{conformer}
A.~Gulati, J.~Qin, C.-C. Chiu, N.~Parmar, Y.~Zhang, J.~Yu, W.~Han, S.~Wang,
  Z.~Zhang, Y.~Wu, and R.~Pang, ``{Conformer: Convolution-augmented Transformer
  for Speech Recognition},'' in \emph{Proc. Interspeech 2020}, 2020, pp.
  5036--5040.

\bibitem{tau19}
\BIBentryALTinterwordspacing
T.~Heittola, A.~Mesaros, and T.~Virtanen, ``{TAU Urban Acoustic Scenes 2019,
  Development dataset},'' Mar. 2019. [Online]. Available:
  \url{https://doi.org/10.5281/zenodo.2589280}
\BIBentrySTDinterwordspacing

\bibitem{specaugment}
\BIBentryALTinterwordspacing
D.~S. Park, W.~Chan, Y.~Zhang, C.-C. Chiu, B.~Zoph, E.~D. Cubuk, and Q.~V. Le,
  ``Specaugment: A simple data augmentation method for automatic speech
  recognition,'' \emph{Interspeech 2019}, Sep 2019. [Online]. Available:
  \url{http://dx.doi.org/10.21437/Interspeech.2019-2680}
\BIBentrySTDinterwordspacing

\bibitem{speechbrain}
M.~Ravanelli, T.~Parcollet, P.~Plantinga, A.~Rouhe, S.~Cornell, L.~Lugosch,
  C.~Subakan, N.~Dawalatabad, A.~Heba, J.~Zhong, J.-C. Chou, S.-L. Yeh, S.-W.
  Fu, C.-F. Liao, E.~Rastorgueva, F.~Grondin, W.~Aris, H.~Na, Y.~Gao, R.~D.
  Mori, and Y.~Bengio, ``{SpeechBrain}: A general-purpose speech toolkit,''
  2021, arXiv:2106.04624.

\end{thebibliography}

\end{document}